\documentclass[proof]{WileyASNA-v1}

\articletype{Article Type}%

\received{}
\revised{}
\accepted{}

\begin{document}

\title{First {\it eROSITA}-{\it TESS} results for M~dwarfs: Mass dependence of the X-ray activity rotation relation and an assessment of sensitivity limits} 
\author[1]{E. Magaudda*}

\author[1,2]{B. Stelzer}

\author[1]{St. Raetz}

\authormark{E. Magaudda \textsc{et al}}

\address[1]{\orgdiv{Institut für Astronomie und Astrophysik}, \orgname{ Eberhard-Karls Universität Tübingen}, \orgaddress{\state{Sand 1, D-72076, Tübingen}, \country{Germany}}}

\address[2]{\orgdiv{INAF-} \orgname{Osservatorio Astronomico di Palermo}, \orgaddress{\state{Piazza Parlamento 1, I-90134, Palermo}, \country{Italy}}}

\corres{*{magaudda@astro.uni-tuebingen.de}}

\abstract{We present a study of the activity-rotation relation for M~dwarf stars, using new X-ray data from the {\it ROentgen Survey with an Imaging Telescope Array (eROSITA)} on board the Russian Spektrum-Roentgen-Gamma mission (SRG), combined with photometric rotation periods from the {\it Transiting Exoplanet Survey Satellite (TESS)}. The stars used in this work are selected from the {\sc superblink} proper motion catalog of nearby M dwarfs. We study the $135$ stars with both a detection in the first {\it eROSITA} survey (eRASS1) and a rotation period measurement from {\it TESS} jointly  with the sample of $197$ {\sc superblink} M~dwarfs re-adapted from our previous work. We fit the activity-rotation relation for stars with rotation periods shorter than $\sim 10$\,d (saturated regime) using three mass bins. The surprising positive slope for stars in our lowest mass bin ($M_{\star} \leq 0.4 {\rm M_\odot}$) is due to a paucity of stars with intermediate rotation periods ($\sim 1-10$\,d), probably caused by fast period evolution. The much higher fraction of eRASS1 detections compared to stars that have also rotation periods from {\it TESS} shows that {\it eROSITA} is also sensitive for slower rotating M~dwarfs that are in the unsaturated regime with periods inaccessible to {\it TESS}.}


\maketitle

\section{Introduction}\label{sec1}
    Solar- and later-type main sequence and giant stars experience magnetic activity in the inner and outer atmosphere.
    The mechanism that powers and enhances the magnetic field in solar-like stars is known to be an $\alpha\Omega-$dynamo, i.e. it operates with combined contributions of the convective motions of the outer stellar envelope ($\alpha$) and the stellar rotation ($\Omega$). 
    As a consequence, cooler regions than the surroundings, so-called dark spots, form in the photosphere, while brighter and hotter structures arise in chromosphere and corona. The latter ones produce copious UV, X-ray, and radio emission. In later-type stars, close and beyond the fully convective transition (SpT $\sim$ M3), the mechanism driving the continuous formation and variation of magnetic fields is not well understood. An indirect measure of this phenomenon is investigating the coronal activity-rotation relation, typically expressed in terms of the X-ray luminosity ($L_{\rm x}$) as a function of the rotational period ($P_{\rm rot}$). 
    \citet{Pallavicini1981} were the first to investigate the coronal X-ray emission as a function of rotational velocity for a wide range of stellar spectral types (O3 to M). Later, \citet{Pizzolato2003} focused their study on late-type main-sequence stars with X-ray data from the {\it ROSAT} satellite and $P_{\rm rot}$ calculated from $v \sin i$ measurements. In this early work the activity-rotation relation presents two different regimes: (1) saturated, where the activity of fast rotators does not depend on rotation, and (2) unsaturated, where the X-ray activity of slowly rotating stars decreases with increasing rotation period. 
        
    In the most recent work from \cite{Magaudda2020} we presented a comprehensive study of the relation between rotation, X-ray activity and age for $\sim 300$ M dwarfs. We homogenized data from the literature  \citep{Wright2011,Wright2018,Wright2016,Stelzer2016,Gonzalez-Alvarez2019} and added in new very sensitive observations from dedicated observations with the X-ray satellites {\it XMM-Newton} and {\it Chandra} and the photometry mission {\it K2} from which we derived rotation periods. We found a significantly steeper slope in the unsaturated regime than previous works for fully convective stars and we confirmed a non-constant X-ray emission level in the saturated regime, as first proposed by \cite{Reiners2014}.   
        
    In this article, we combine our previous results on the X-ray activity-rotation relation \citep{Magaudda2020} with those obtained from new observations with the {\it extended ROentgen survey with an Imaging Telescope Array} \citep[eROSITA;][]{Predehl2021} on the Russian Spektrum Roentgen-Gamma (SRG)\footnote{\url{http://srg.iki.rssi.ru/?page_id=676&lang=en}} mission and the {\it Transiting Exoplanet Survey Satellite} \citep[TESS;][]{Ricker2014}. 

    We use the {\it eROSITA} and {\it TESS} sample compiled by \cite{Magauddasubm_2021}, on the basis of the {\sc superblink} proper motion survey by \cite{Lepine2011} (LG11, $\sim 9000$ M~dwarfs), that we characterized with {\it Gaia}-DR2 data. We refer to \cite{Magauddasubm_2021} for more information on the derivation of our final catalog of LG11-{\it Gaia} M~dwarfs detected with {\it eROSITA} and with {\it TESS} observations.
    More details on the sample used for the analysis of the activity-rotation relation treated in this work are also given in Sect.~\ref{sec:sample}, and 
    we present the results in Sect.~\ref{sec:discussion}, with an investigation of the possible observational biases related to them.
    In Sect.~\ref{sec:conclusions} we summarize our conclusions and give an outlook to future studies in this field. 

\section{Sample}\label{sec:sample}
We selected M~dwarfs from the {\sc superblink} proper motion catalog of nearby M dwarfs from \citet{Lepine2011} choosing only stars with complete photometry from the second data release of the {\it Gaia} mission \citep[\textit{Gaia} DR2,][]{GaiaColl2018} and distance from \citet{BailerJones2018} (hereafter LG11-{\it Gaia}).
In particular, we study the X-ray emission of our LG11-{\it Gaia} sample during the first {\it eROSITA} All-Sky survey (eRASS1). 
We use the preliminary catalog produced by the {\it eROSITA} consortium that is based on data collected from Dec 2018 to June 2019. We retrieve $P_{\rm rot}$ from {\it TESS} light curves with $2$-min cadence. A complete discussion of the {\it eROSITA}/{\it TESS} source identification and data analysis is found in \cite{Magauddasubm_2021}. 

The sample used in this work includes the LG11-{\it Gaia} M~dwarfs detected with {\it eROSITA} and with reliable $P_{\rm rot}$ from {\it TESS} light curves. We refer to it as the LG11-{\it Gaia}/eRASS1/{\it TESS} sample. The relations we used to calculate stellar parameters \citep{Mann2015,Mann_2016} are valid for stars with $4.6<M_{\rm Ks}<9.8$ and $0.1<R_{\star}/R_{\odot}<0.7$, therefore we keep only the sub-sample for which these conditions apply and we call it  the `validated' LG11-{\it Gaia}/eRASS1/{\it TESS} catalog. 

Next to the `validated' LG11-{\it Gaia}/eRASS1/{\it TESS} sample we use the {\it MagauddaGaia20} catalog, that was selected from the one used in \cite{Magaudda2020} where for consistency we consider only X-ray detected sources (i.e. we disregard all upper limits) with complete {\it Gaia}-DR2 and in the validity range of the relations from \cite{Mann2015,Mann_2016}, amounting to $197$ stars.

\section{Results \& Discussion}\label{sec:discussion}

\subsection{The X-ray activity-rotation relation}\label{sec:act_rot}
\begin{center}
    \begin{table}[t]
    \centering
    \caption{Best fit parameter results.\label{tab:fit_res}}%
    \begin{tabular*}{22pc}{@{\extracolsep\fill}lccc@{\extracolsep\fill}}%
        \toprule
        \textbf{Mass Bin} & \textbf{Nstars} &\textbf{$\beta$} & \textbf{Intercept} \\
        &&& [erg/s]\\\midrule
        Full Sample&242&+0.12$\pm$0.06&28.78$\pm$0.03\\
        $M_{\star}/M_{\odot} > 0.6$ & 56 & $-0.05\pm$0.09 & 29.36$\pm$0.03 \\
        $0.4 \leq M_{\star}/M_{\odot} \leq 0.6$& 120 &  $-0.19\pm$0.06 & 28.90$\pm$0.03 \\
        $M_{\star}/M_{\odot} < 0.4$& 72 & +0.24$\pm$0.07 & 28.18$\pm$0.04 \\
        \bottomrule
    \end{tabular*}
    \end{table}
\end{center}
New data from the {\it eROSITA} and {\it TESS} satellites provide unprecedented statistics of fast rotating M~dwarfs located in the saturated regime of the X-ray activity-rotation relation. In Fig.~8 from \citet{Magauddasubm_2021} we can appreciate the capability of {\it eROSITA} in  detecting stars. In fact the amount of X-ray detected stars we present in \citet{Magauddasubm_2021} is the largest to date for M~dwarfs in the saturated regime ($P_{\rm rot}\lesssim 10$\,d), and extends to the lowest X-ray luminosities measured for fast rotators so far. The absence of stars with higher $P_{\rm rot}$ is explained by the duration of the {\it TESS} campaigns ($\sim 27$\,d) which impedes the detection of periods of slower rotators, and consequently a study of the non-saturated region of the relation. Thus, our new analysis is focused on the saturated regime, where we added the new `validated' LG11-{\it Gaia}/eRASS1/{\it TESS} sample to {\it MagauddaGaia20}. 

In \cite{Magaudda2020} we provided a statistical analysis of both the saturated and the unsaturated regimes. Therein we analyzed the activity-rotation relation first for the whole sample and then derived it within three mass bins, (1) $M_{\star}/M_{\odot}>0.6$, (2) $0.4\leq M_{\star}/M_{\odot}\leq0.6$, and (3) $M_{\star}/M_{\odot}<0.4$. 

Following that work, we analyze here the $L_{\rm x}-P_{\rm rot}$ relation in the saturated regime of the full new sample, i.e. `validated' LG11-{\it Gaia}/eRASS1/{\it TESS} plus {\it MagauddaGaia20} stars, and then in individual mass bins.

\begin{figure}
    \centering
    \includegraphics[width=0.50\textwidth]{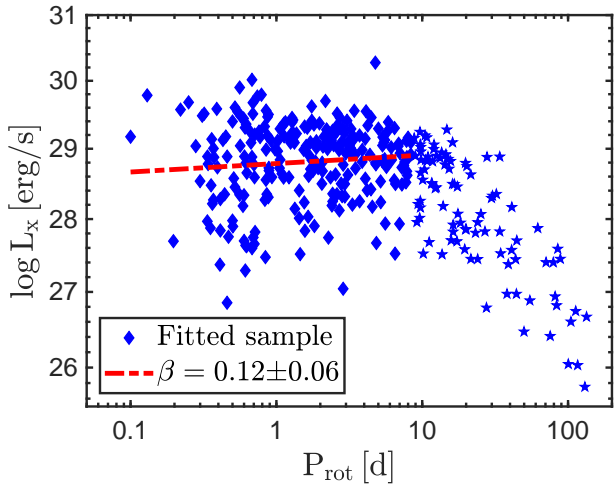}
    \caption{The X-ray activity-rotation relation for the sample of this work, where new X-ray data from {\it eROSITA} and photometric rotation periods from {\it TESS} were added to the sample presented by \citet{Magaudda2020}. The power law fit in the saturated regime, $P_{\rm rot} \leq 8.5$\,d  (red dashed line), and its slope $\beta$ are shown.}
    \label{fig:LxProt_fullsample}
\end{figure}

In the analysis of our new sample across the whole `validated' mass range we use the transition from the saturated to the non-saturated regime found in \citet{Magaudda2020} for the full mass sample ($P_{\rm sat} = 8.5$\,d) to define the saturated region for the sample.
We performed a power law fit to the $L_{\rm x}-P_{\rm rot}$ data for the full new sample. The best fit parameters are listed in Table~\ref{tab:fit_res} and the value for the slope is also shown in Fig.~\ref{fig:LxProt_fullsample}. Surprisingly, a positive slope is found, i.e. $L_{\rm x}$ increases with increasing $P_{\rm rot}$, contrary to all previous studies of the saturated regime that found $L_{\rm x} \approx\rm const$ or a slightly decreasing slope \citep{Magaudda2020,Reiners2014}.

\begin{figure*}[htbp]
	\begin{center}
	\parbox{18cm}{
			\parbox{6cm}{\includegraphics[width=0.35\textwidth]{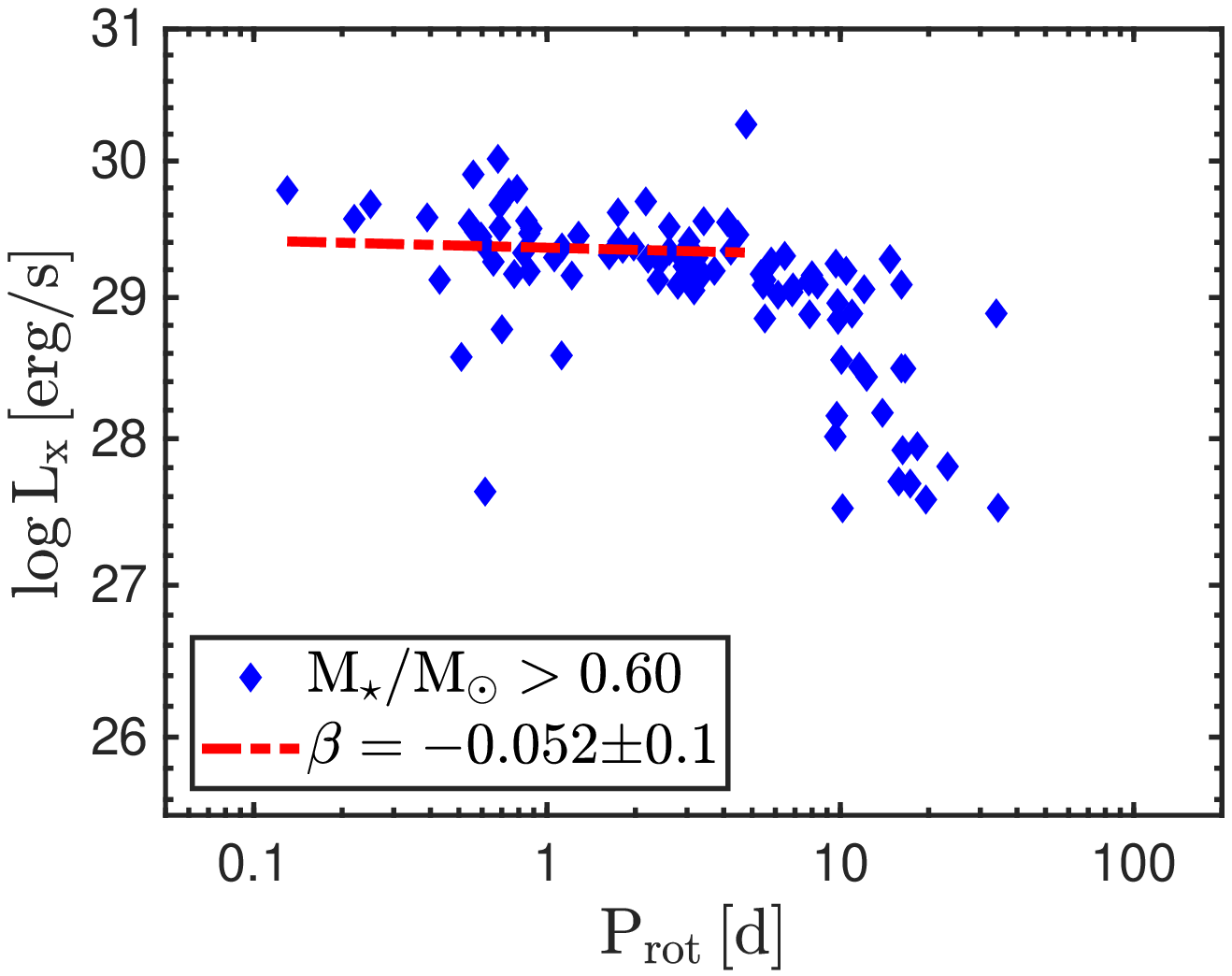}}
			\parbox{6cm}{\includegraphics[width=0.35\textwidth]{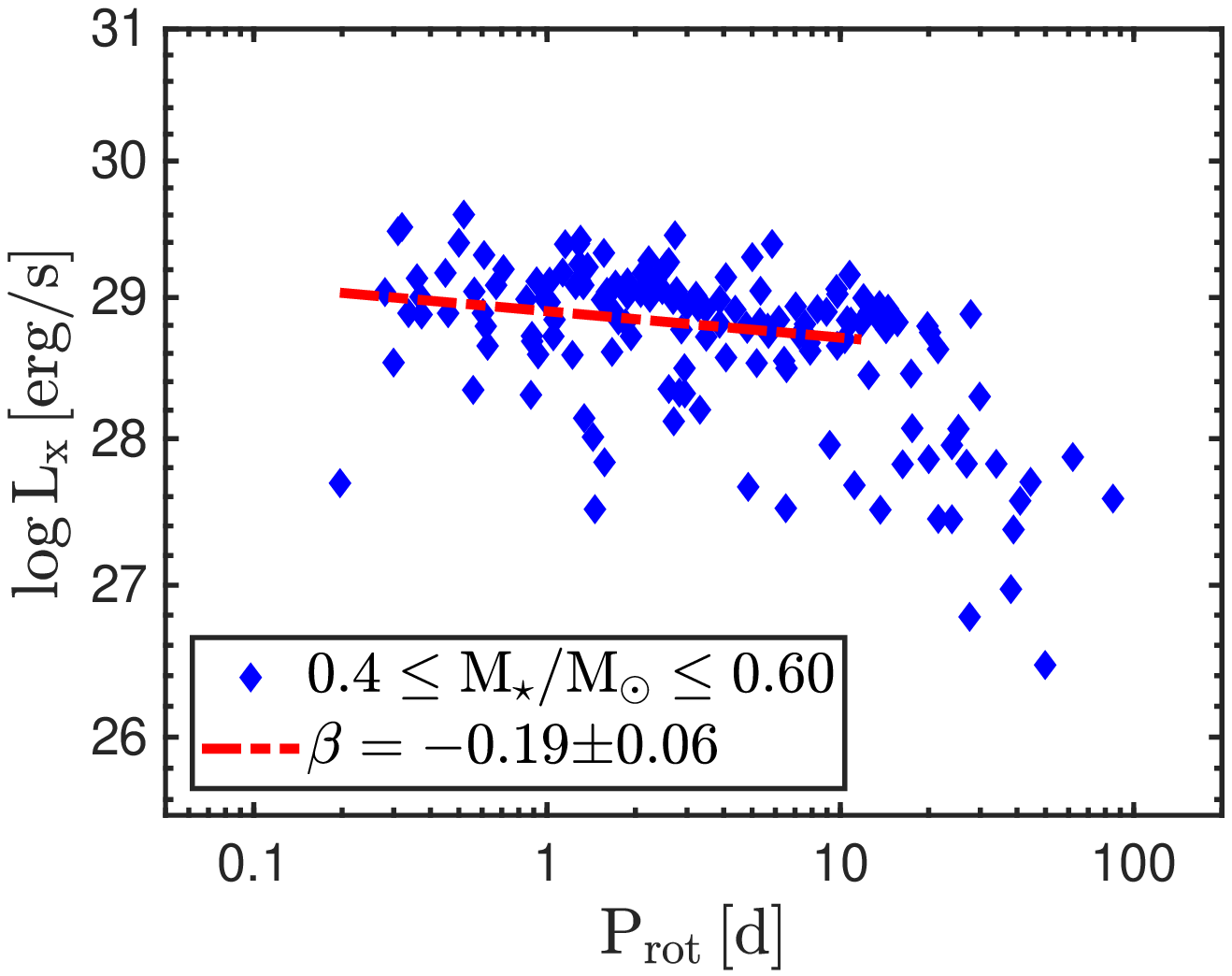}}
			\parbox{6cm}{\includegraphics[width=0.35\textwidth]{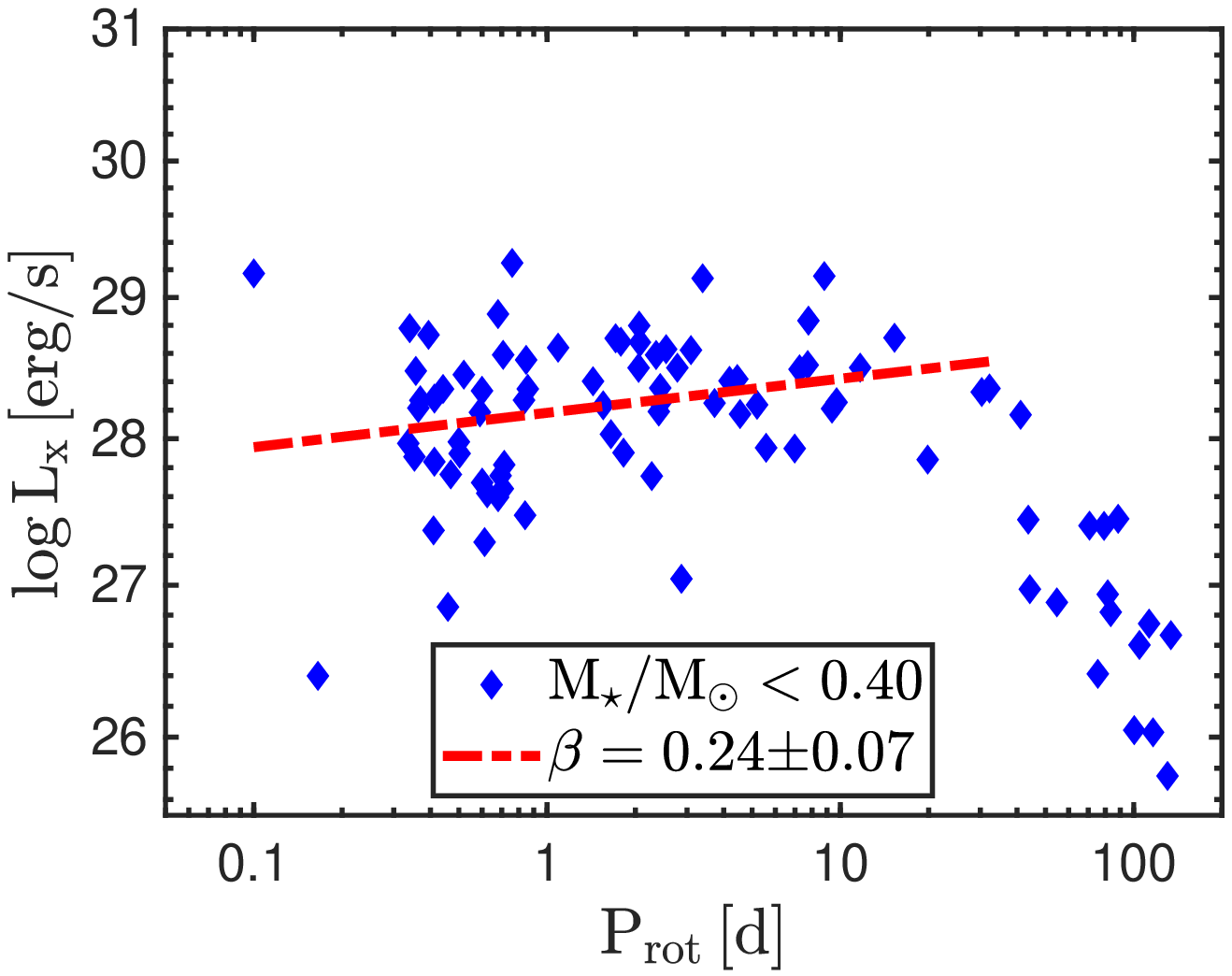}}
			}
			\parbox{18cm}{
			\parbox{12cm}{\caption{$L_{\rm x}-P_{\rm rot}$ relation and power law fits in the stellar mass bins defined in Sect.~\ref{sec:act_rot}.\label{fig:LxProt_massbins}}}
			\parbox{6cm}{\includegraphics[width=0.35\textwidth]{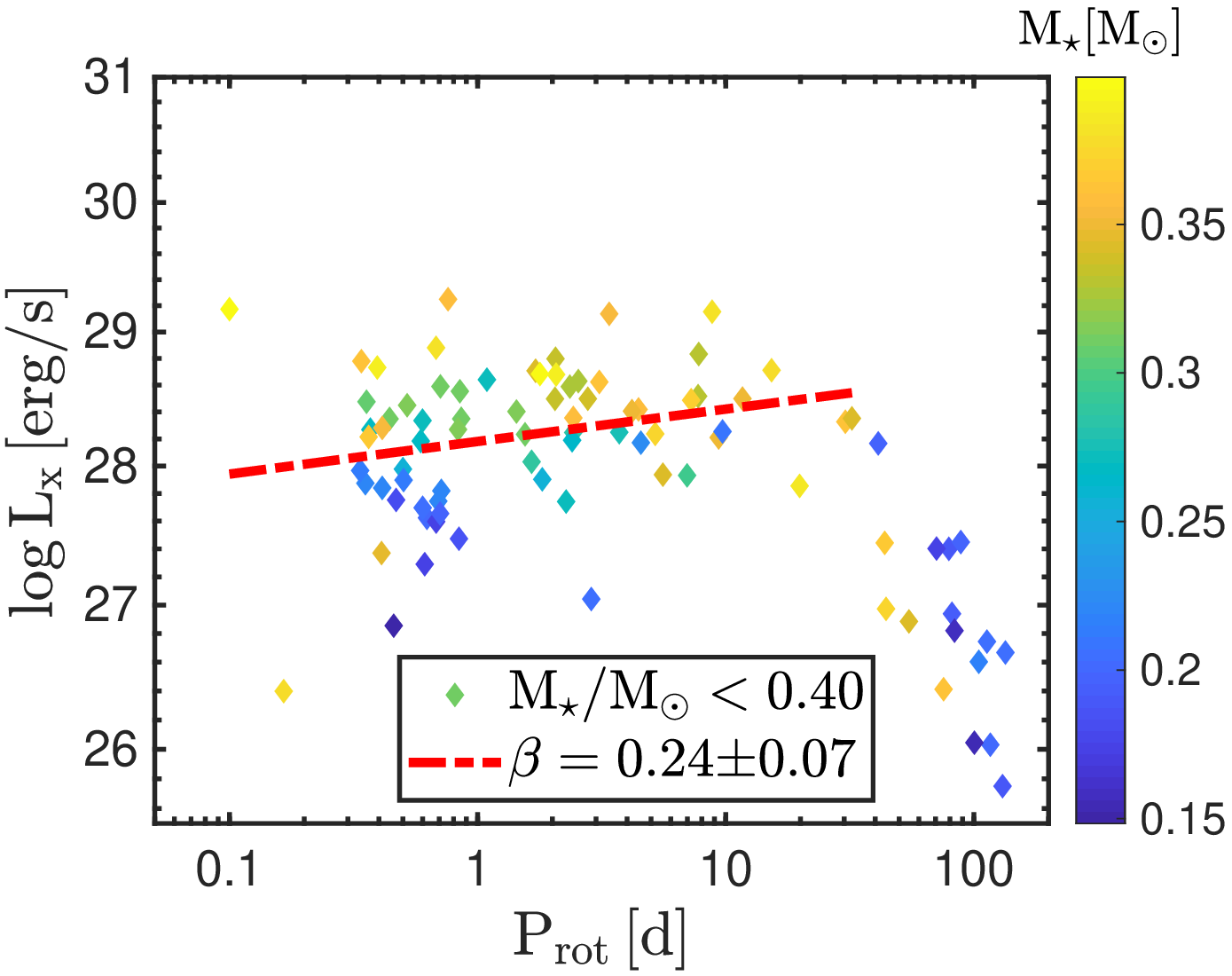}}
			}
	
	\end{center}
\end{figure*}

To examine potential observational biases, we analyzed the relation in three different mass bins, each with its own maximum period for the saturated regime ($P_{\rm rot,sat}$) from the analysis performed by \citet{Magaudda2020}.
In particular, $P_{\rm rot,sat}~(M_{\star}/M_{\odot}>0.6)=5.2\pm 0.7$\,d, $P_{\rm rot,sat}~(0.4\leq M_{\star}/M_{\odot}\leq0.6)=11.8\pm 2.0$\,d and $P_{\rm rot,sat}~(M_{\star}/M_{\odot}<0.4)=33.7\pm 4.5$\,d. In Fig.~\ref{fig:LxProt_massbins} we show the results of the power law fits for the three mass bins, and the best fit parameters are, as before, listed in Table~\ref{tab:fit_res}. It is evident that the slope is positive only in the lowest mass bin. In this bin $L_{\rm x}$ covers a wide range, going from the minimum $L_{\rm x}-$level detected for M~dwarfs with very short $P_{\rm rot}$ measurements ($P_{\rm rot}\leq 0.5$\,d) to almost its maximum value. 
In the bottom panel of Fig.~\ref{fig:LxProt_massbins} we present the $L_{\rm x}-P_{\rm rot}$ relation for the lowest mass bin with a $M_{\star}-$color code.
Here we can see that the period distribution of very low mass stars ($M_{\star}/M_{\odot}\leq 0.2$) is composed of stars with very short or very long $P_{\rm rot}$ (in the saturated and unsaturated regime, respectively). Observational biases are strong as already mentioned in \cite{Magaudda2020} and \cite{Magauddasubm_2021}: very long rotation periods are taken from \citet{Wright2018} who selected data from the MEarth project to study the slow rotator regime. {\it TESS} observations provide periods shorter than $\sim 15$\,d because {\it TESS} monitors a given field for only about a month, and we considered reliable all $P_{\rm rot}$ shorter than about half the duration of a sector light curve. However, these biases can not explain the absence of stars with $P_{\rm rot}\approx 1-10$\,d that is present in these very low mass stars ($M_{\star}/M_{\odot}<0.2$). The paucity of very low mass M~dwarfs with intermediate rotation periods is probably caused by fast angular momentum loss that carries the stars in little time from being fast to slow rotators \citep{Newton2017}. Hence, in this scenario it is intrinsically rare to observe late M~dwarfs with rotation periods of $\approx1-10$\,d.

In the highest and the intermediate mass bins the $L_{\rm x}-P_{\rm rot}$ relation shows a marginally negative slope ($\beta_{high} = -0.05\pm 0.09$ and $\beta_{int} = -0.19\pm 0.06$) more in line with previous estimates, although slightly smaller than what we found in \citet{Magaudda2020}, where $\beta_{high} = -0.17\pm 0.14$ and $\beta_{int} = -0.39\pm 0.13$. 

\subsection{{\it eROSITA} detection sensitivity}\label{sec:sens_er}

\begin{figure*}[htbp]
	\begin{center}
	\parbox{18cm}{
		\parbox{6cm}{\includegraphics[width=0.35\textwidth]{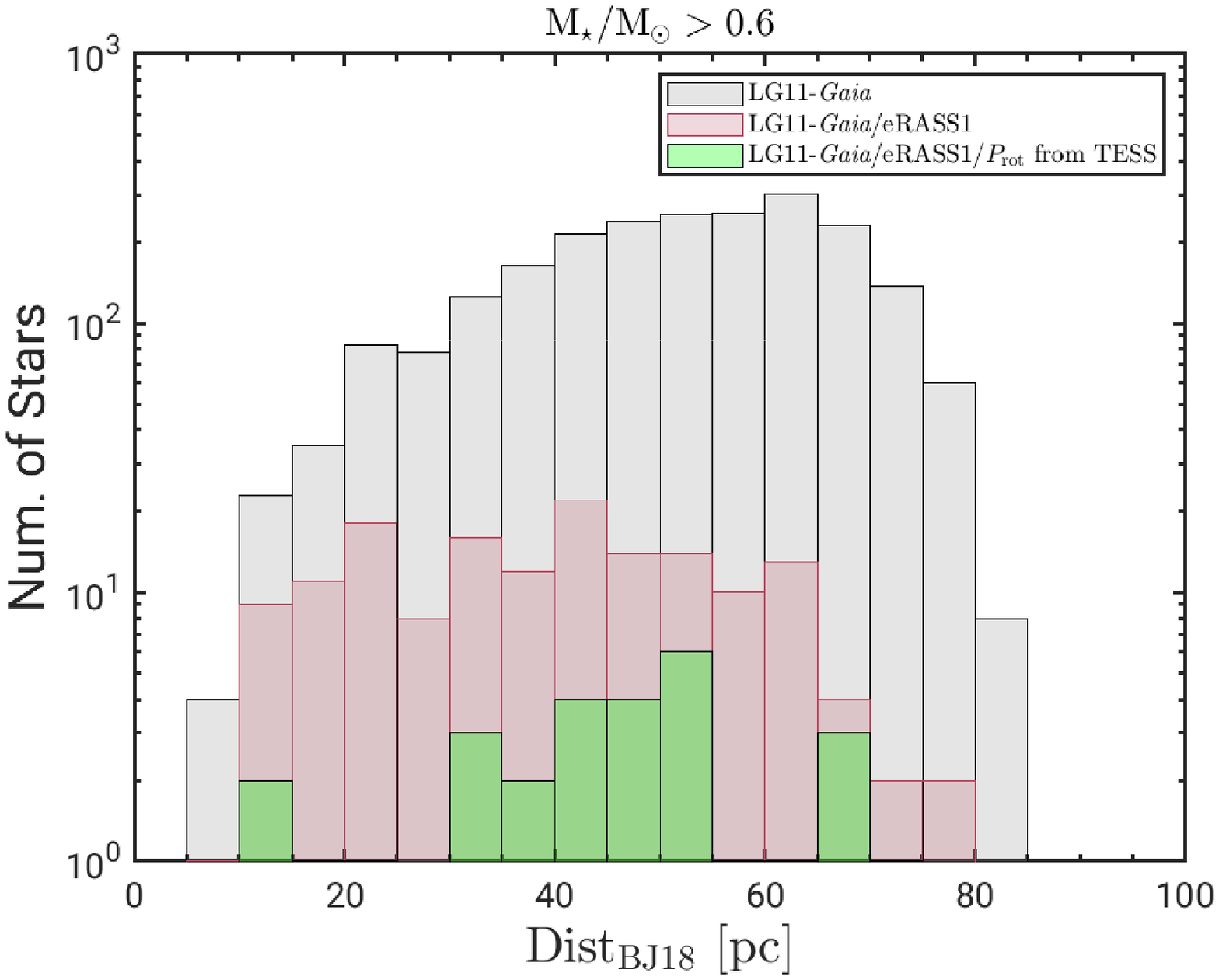}}
		\parbox{6cm}{\includegraphics[width=0.35\textwidth]{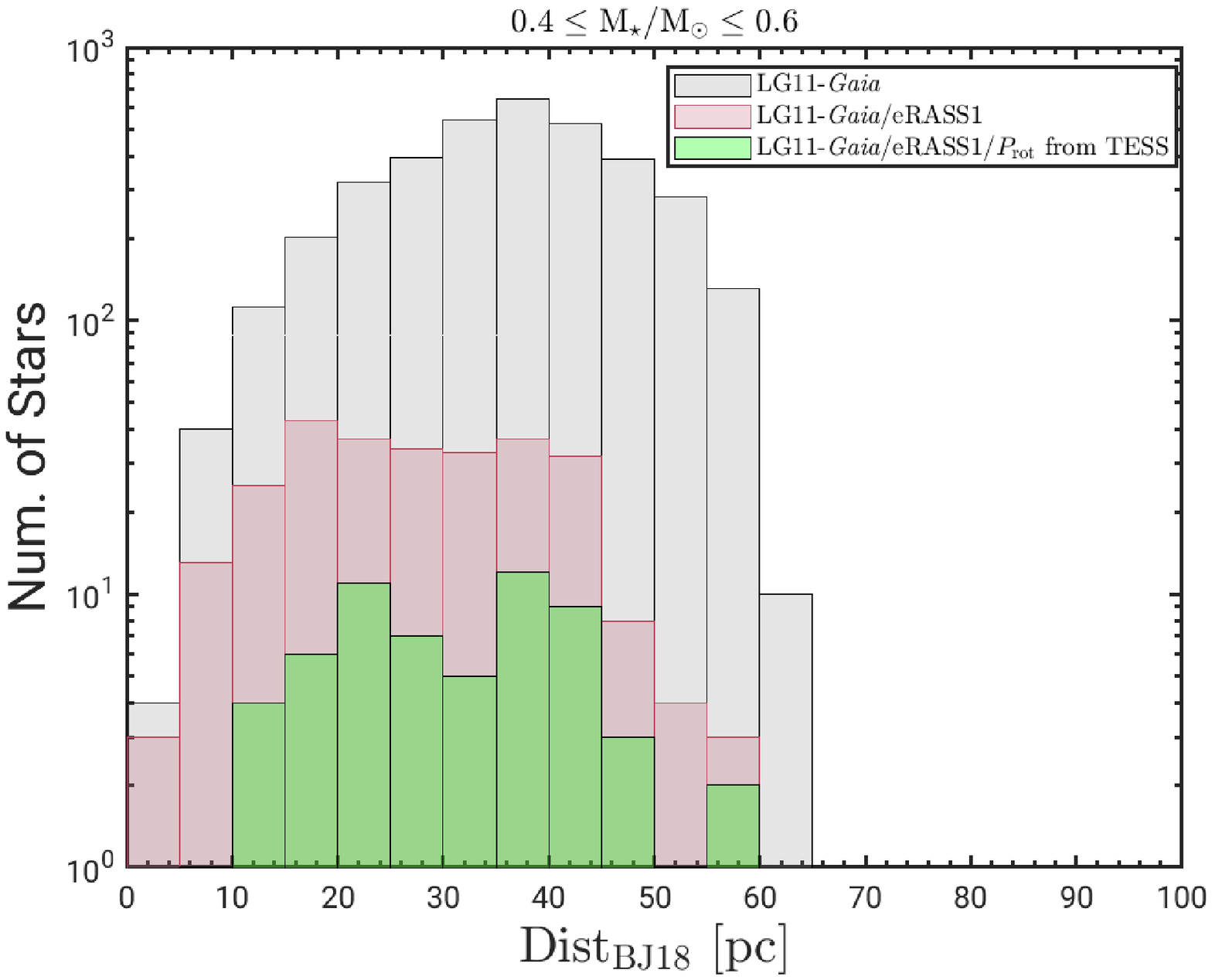}}	\parbox{6cm}{\includegraphics[width=0.35\textwidth]{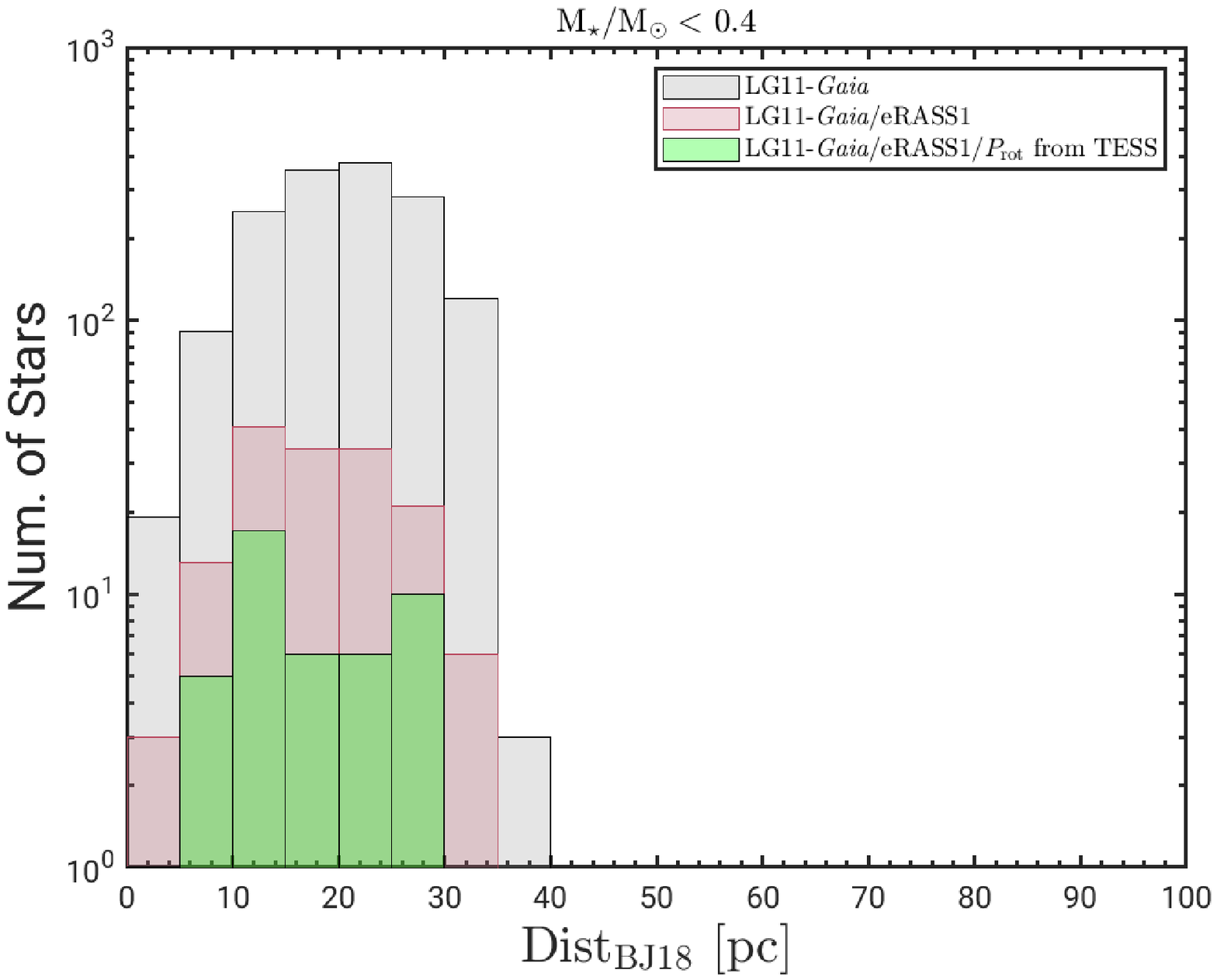}}

	}
	\caption{Distance distributions in the three stellar mass bins defined in Sect.~\ref{sec:sens_er} for the full LG11-{\it Gaia} sample (in gray) in comparison with the M~dwarf sub-sample detected only with {\it eROSITA} during eRASS1 (in red) and the one detected with {\it eROSITA} and with reliable $P_{\rm rot}$ from {\it TESS} light curves (in green). In the highest mass bin there are four stars with $Dist_{\rm BJ18}>100$\,pc that we do not show to have a better visualization of the distance distribution.}
	\label{fig:hist_dist_massbins}
	\end{center}
\end{figure*}

\begin{figure}
    \centering
    \includegraphics[width=0.50\textwidth]{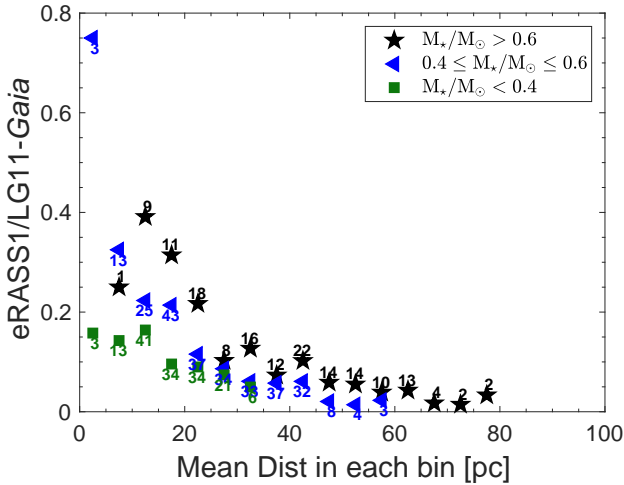}
    \caption{Ratio between the red and gray samples of the histograms in Fig.~\ref{fig:hist_dist_massbins}, calculated for each mass bin. We indicate the number of stars in each bin for the eRASS1 detections, i.e. the LG11-{\it Gaia}/eRASS1 sample.}
    \label{fig:LG11_eR_ratio}
\end{figure}

\begin{figure}[htbp]
	\begin{center}
	\includegraphics[width=0.50\textwidth]{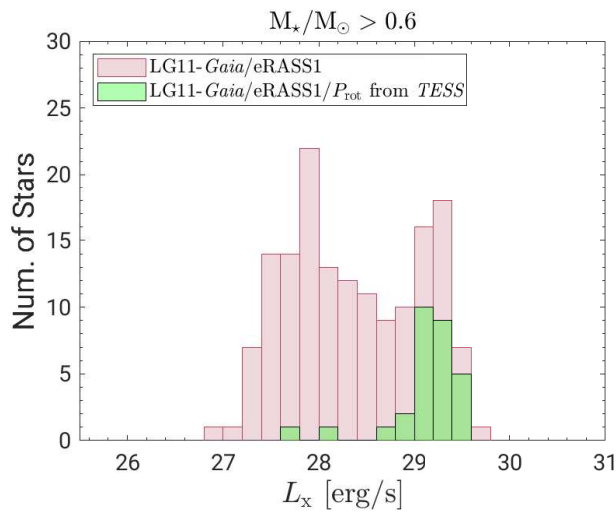}
	\caption{X-ray luminosity distribution for the highest mass bin of LG11-{\it Gaia} stars detected in eRASS1 compared to that of the subsample with reliable $P_{\rm rot}$.}
	\label{fig:hist_Lx}
	\end{center}
\end{figure}
	
\begin{figure}
    \centering
    \includegraphics[width=0.50\textwidth]{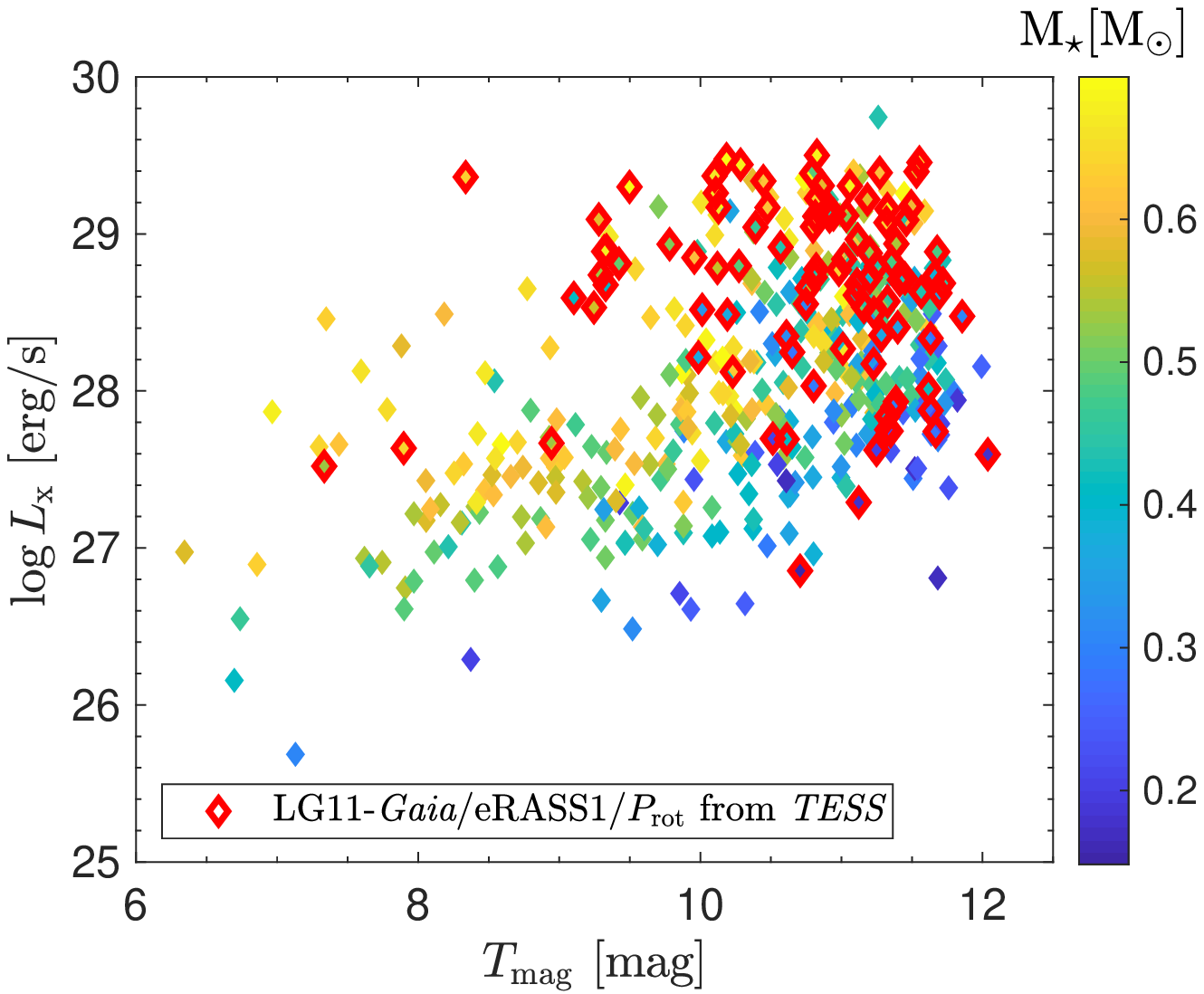}    
    \caption{The relation between $L_{\rm x}$ and {\it TESS} magnitude for the LG11-{\it Gaia}/eRASS1 sample shown with a mass color code and the same relation for those detections with reliable $P_{\rm rot}$ (open red diamonds).}
    \label{fig:LxTmag}
\end{figure}

Through the study of the activity-rotation relation in Sect.~\ref{sec:act_rot} we encountered observational biases that call for a thorough investigation of the {\it eROSITA} detection statistics for the different mass bins.

The LG11-{\it Gaia} sample, with stellar parameters that are within the validation range of the relations from \citet{Mann2015,Mann_2016}, counts $7319$ stars with complete {\it Gaia}-DR2 data, i.e. photometry and distance. 
Eight percent of this sample is identified with eRASS1 sources, and $\sim~6$\,\% is detected  in eRASS1 and has a {\it TESS} 2-min light curve. We extracted reliable $P_{\rm rot}$ from {\it TESS} light curves for $135$ stars. This corresponds to $\sim~2\,\%$ of the whole LG11-{\it Gaia} sample and $23$\,\% of the eRASS1 detections. Thus, as we mentioned in \cite{Magauddasubm_2021}, monitoring coronal X-ray emission with {\it eROSITA} is more efficient in identifying stellar magnetic activity than photometric starspots with {\it TESS}. 
 
To investigate the {\it eROSITA} detection statistics we explored the distance ($Dist_{\rm BJ18}$) and mass distributions of the different samples, and how they relate to the detection efficiency of the instrument.
 
 We separated the LG11-{\it Gaia} sample into the three $M_{\star}-$bins used in Sect.~\ref{sec:act_rot}. In Fig.~\ref{fig:hist_dist_massbins} we distinguish the complete LG11-{\it Gaia} sample (gray histogram) from those sources detected with {\it eROSITA} and those having both an {\it eROSITA} detection and reliable {\it TESS} $P_{\rm rot}$ (red and green histograms, respectively). The LG11-{\it Gaia} catalog consists of nearby ($Dist_{\rm BJ18} < 100$\,pc) and bright ($J < 10$\,mag) M~stars according to the selection criteria adopted by \citet{Lepine2011}. In Fig.~\ref{fig:hist_dist_massbins} the maximum distance is indeed $\sim 100$\,pc with $0.05$\,\% outliers with distances up to $\sim250$\,pc for massive M dwarfs and it decreases for lower mass stars that are fainter.

 From Fig.~\ref{fig:hist_dist_massbins} we see that the eRASS1 distance limit is only slightly lower than that of the LG11-{\it Gaia} catalog as a whole, but the fraction of eRASS1 detections decreases with distance. This is quantified in Fig.~\ref{fig:LG11_eR_ratio} where we present the fraction of X-ray detected LG11-{\it Gaia} M~dwarfs vs distance  separately for the three $M_{\star}-$bins. In other words, we show the ratio between the red and the gray histograms of Fig.~\ref{fig:hist_dist_massbins}. As a measure for the  statistics we indicate the number of X-ray detected stars in each distance bin attached to the plotting symbol. From this representation it is evident that at a given distance the fraction of eRASS1 detected stars decreases with decreasing mass. Apart from one outlier in the intermediate mass bin, the detection fraction for lower distances reaches up to $\approx30\,\%$ while it is $<10$\,\%  for distances beyond $\approx 20$\,pc.
 
 Comparing the eRASS1 sample to the subsample that also has reliable rotation periods from TESS (red and green histograms in Fig.~\ref{fig:hist_dist_massbins}) we can see that there is no clear difference in their distance limit, but the amount of detected stars in the eRASS1 sample is larger than the one in the subsample with reliable $P_{\rm rot}$, especially for the highest mass bin. In Fig.~\ref{fig:hist_Lx} we show the $L_{\rm x}$ distribution for M~dwarfs with $M_{\star}/M_{\odot}>0.6$ for the eRASS1 sample with the one for eRASS1 stars with reliable $P_{\rm rot}$ (same color code as in Fig.~\ref{fig:hist_dist_massbins}). We see that periods have been detected only in the stars with the highest $\log L_{\rm x}$ ($\gtrapprox29.0$\,erg/s). 
 According to the activity-rotation relation for the highest mass bin (top left panel in Fig.~\ref{fig:LxProt_massbins}) stars with $L_{\rm x}\lesssim29$\,erg/s are expected to have $P_{\rm rot}\gtrsim 10$\,d.
 We have verified that the situation is similar for the intermediate-mass bin. Therefore, except for the lowest-mass M~dwarfs, eRASS1 can detect X-ray emission from unsaturated stars, for which we can not detect rotation periods with {\it TESS}. 

To further investigate how the capability of {\it TESS} for detecting rotation periods depends on other parameters we show in Fig.~\ref{fig:LxTmag} a diagram of X-ray luminosity versus {\it TESS} magnitude\footnote{We have calculated $T_{\rm mag}$ with the conversion from {\it Gaia} photometry provided by \cite{Stassun2019}.} ($T_{\rm mag}$). Stars for which we were able to derive $P_{\rm rot}$ values are highlighted in red. We show the rest of the eRASS1 sample with a mass color code. From this representation we see (1) that rotation periods are mostly found for stars with high $L_{\rm x}$-level as already shown in Fig.~\ref{fig:hist_Lx} and (2) that very few rotation periods are detected on optically bright stars ($T_{\rm mag}\lesssim 9$\, mag). This is surprising at first sight, as rotational modulation should be more easy to identify in stars with brighter $T$ magnitude due to the higher photometric precision.

We first focus on early M dwarfs (yellow and green in Fig.~\ref{fig:LxTmag}). It is clear that there is a lack of optically bright ($T_{\rm mag}\lesssim 9$\, mag) stars with $\log{L_{\rm x}}\,{\rm [erg/s]} \gtrsim 28$. The optically brightest stars sample the shortest distances. 
Thus, by the absence of stars in the upper left of Fig.~\ref{fig:LxTmag} we conclude that the sample is devoid of very nearby early M dwarfs in the saturated regime. This can be explained if at least for the two higher-mass bins, the unsaturated regime holds many more stars than the saturated regime (represented by the yellow and green objects at bright $T_{\rm mag}$ but low $L_{\rm x}$). In fact, in the large sample of M~dwarfs analyzed by \cite{McQuillan2013,McQuillan2014} and based on three years of {\it Kepler} data the period distribution shows an upper envelope that increases for decreasing masses. In particular, stars in the mass range studied in this work are largely detected in the saturated regime ($P_{\rm rot} \textcolor{blue}{\lesssim} 10$\,d).
For the lowest mass stars ($M_* \leq 0.4\,{\rm M_\odot}$) the $T$ magnitude distribution is similar than that of the higher-mass stars but shifted to lower $L_{\rm x}$.  However, here the absence of period detections in the optically brighter (more nearby) stars is due to the fact that these have low $L_{\rm x}$ values, that is they are unsaturated and their periods are undetectably long for {\it TESS}. Optically fainter --  that is more distant -- stars at the same (low) X-ray luminosity are entirely absent from the sample (see the void of objects in the lower right of Fig.~\ref{fig:LxTmag}) because they are beyond the eRASS1 flux limit.

\section{Conclusions}\label{sec:conclusions}
We investigated the X-ray activity-rotation relation for M~dwarfs combining the sample from \citet{Magaudda2020} with new X-ray data from {\it eROSITA} and photometric rotation periods extracted from {\it TESS} light curves. We performed a power law fit of the saturated regime in three mass bins, finding that stars with $M_{\star}<0.4\,M_{\odot}$ show a positive slope due to the paucity of mid-to-late M~dwarfs with intermediate rotation periods  \citep{Newton2017}. 
Past studies \citep[e.g.][]{Prosser1996, Jeffries2011} proposed a so-called `super-saturation' regime where at very fast rotation rates (and low Rossby number) the coronal activity is reduced. 
The rising slope that we find in the right panels of Fig.~\ref{fig:LxProt_massbins} is driven by the lowest mass stars ($M_* \lesssim 0.2\,{\rm M_\odot}$). 
However, we can not ascertain whether some of them represent supersaturated downward outliers because of the absence of stars with the same mass at intermediate periods ($\approx 1-10$\,d) mentioned above.

In the course of our investigation of sensitivity limits we explored the mass and distance dependencies finding that: (1) at a given distance the fraction of eRASS1 detections decreases with decreasing stellar mass, a result of the mass-dependence of $L_{\rm x}$ and the eRASS1 sensitivity limit.
(2) {\it eROSITA} is sensitive to detect stars with larger rotation periods that are not detectable with {\it TESS}, (3) for early M dwarfs the saturated regime is very poorly sampled with a very small sky volume indicating that in this mass range the majority of stars are unsaturated, i.e. they are already in an evolved phase of their spin-down history consistent with {\it Kepler} $P_{\rm rot}-$distributions.

Clearly, several observational biases mix with intrinsic distributions of X-ray activity and rotation periods. Here we have presented a first assessment of such effects that require a more detailed study in the future.

{\footnotesize
\section*{Acknowledgments}

    EM is supported by the Bundesministerium f\"{u}r Wirtschaft und Energie through the Deutsches Zentrum f\"{u}r Luft- und Raumfahrt e.V. (DLR) under grant number FKZ 50 OR 1808.
        
    This work is based on data from eROSITA, the soft X-ray instrument aboard SRG, a joint Russian-German science mission supported by the Russian Space Agency (Roskosmos), in the interests of the Russian Academy of Sciences represented by its Space Research Institute (IKI), and the Deutsches Zentrum für Luft- und Raumfahrt (DLR). The SRG spacecraft was built by Lavochkin Association (NPOL) and its subcontractors, and is operated by NPOL with support from the Max Planck Institute for Extraterrestrial Physics (MPE).

    This paper includes data collected with the TESS mission, obtained from the MAST data archive at the Space Telescope Science Institute (STScI). Funding for the TESS mission is provided by the NASA Explorer Program. STScI is operated by the Association of Universities for Research in Astronomy, Inc., under NASA contract NAS 5–26555.
}
    
\bibliography{bibliography}%
\end{document}